\documentclass{aa}
\usepackage{graphics}
\input psfig.sty
\newcommand{\ltsima} {$\; \buildrel < \over \sim \;$}
\newcommand{\gtsima} {$\; \buildrel > \over \sim \;$}
\newcommand{\lta} {\lower.5ex\hbox{\ltsima}}
\newcommand{\gta} {\lower.5ex\hbox{\gtsima}}

\begin{document}

\thesaurus{03(11.01.2; 11.02.1; 11.10.1; 11.14.1)} 

\title{Does the unification of BL Lac and FR I radio galaxies require 
jet velocity structures?}

\author{M. Chiaberge \inst{1}, A. Celotti \inst{1}, A. Capetti
\inst{2} \and G. Ghisellini \inst{3}}

\offprints{M. Chiaberge, chiab@sissa.it}

\institute{SISSA/ISAS, Via Beirut 2-4, I-34014 Trieste, Italy \and
Osservatorio Astronomico di Torino, Strada Osservatorio 20, I-10025
Pino Torinese, Italy \and Osservatorio Astronomico di Brera, Via
Bianchi 46, I-23807 Merate (LC), Italy}

\date{Received ...; accepted ...}

\titlerunning{FR~I - BL Lacs unification}
\authorrunning{Chiaberge et al.}   \maketitle

\begin{abstract}

We explore the viability of the unification of BL Lacs and FR~I radio
galaxies by comparing the core emission of radio galaxies  
with those of BL Lacs of similar extended radio power,
taking advantage of the newly measured optical nuclear
luminosity of FR~I sources. 
The spectral properties of complete samples
are also studied in the radio-optical luminosity plane: starting from
the Spectral Energy Distribution (SED) of BL Lacs, we calculate the
predicted luminosity of FR~I nuclei in the frame of a simple one--zone
model, by properly taking into account the relativistic transformations.
We find that the bulk Lorentz factors required by the spread in
the observed luminosities in all bands are significantly smaller than
those implied by other, both observational and theoretical, considerations.
This discrepancy is also reflected in the fact that
FR~I nuclei are over--luminous by a factor of 10-10$^4$, with
respect to the predictions, both in the radio
and in the optical band.

In order to reconcile these results with the unification scheme,
velocity structures in the jet are suggested, where a fast spine is
surrounded by a slow (but still relativistic) layer so that the
emission at different angles is dominated by different velocity
components: the fast one dominates the emission in BL Lacs while the 
slow layer dominates the emission in misaligned objects. 
Furthermore for the lowest luminosity BL
Lacs it has to be also postulated that their beaming factor in the
radio band is lower than in the optical (and X--ray), as would result
from deceleration of the jet.

The self--consistency of the unification model therefore requires that
both intrinsic differences in the SED and different beaming properties
play a substantial role in characterizing the phenomenology of these
sources.


\end{abstract}

\section{Introduction}

Unification models adduce the main differences between the observed
properties of different classes of AGNs to the anisotropy of the
radiation emitted by the active nucleus (see Antonucci \cite{anto} and
Urry \& Padovani \cite{urrypad} for reviews).  In particular, for low
luminosity radio-loud objects, namely BL Lacs and FR~I radio galaxies
(Fanaroff \& Riley \cite{fr74}), it is believed that this effect is mainly
due to relativistic beaming.
In fact, there is growing evidence that obscuration does not play a 
significant role in these objects, contrary to other classes of AGNs.
This is indicated by optical (Chiaberge et al. \cite{paperI} hereafter Paper~I),
radio (Henkel et al. \cite{henkel}), and X-ray information
(e.g. Fabbiano et al. \cite{fabbiano}, Worral and Birkinshaw \cite{wb94},
Trussoni et al. \cite{edo}).
Within this scenario, the emission from the inner regions of
a relativistic jet dominates the observed radiation in BL Lacs, while
in FR~I, whose jet is
observed at larger angles with respect to the line of sight,
this component is strongly debeamed.  Evidence
for this unification scheme includes the power and morphology of the
extended radio emission of BL Lacs (e.g. Antonucci \& Ulvestad
\cite{antoulve}, Kollgaard et al.  \cite{koll92}, Murphy et al. 
\cite{murphy}) and the properties of their host galaxies
(e.g. Ulrich \cite{ulrich89}, Stickel et al. \cite{stickel91}, Urry et
al.  \cite{urry99}), which are similar to those of FR~I.  
Furthermore, there is a quantitative agreement among the
amount of beaming required by different observational properties
(e.g. Ghisellini et al. \cite{gg93}), the number densities and
luminosity functions of the parent and beamed populations in different
bands (e.g. Urry \& Padovani \cite{urrypad}, Celotti et
al. \cite{celo93}) and the comparison of the radio core emission of
beamed and unbeamed objects with similar total radio power
(Kollgaard et al. \cite{koll96}).

Despite this global agreement, it should be stressed that beaming
factors inferred from the broad band spectral
properties of blazars, more specifically superluminal motions,
transparency to the $\gamma$--ray emission, shape of the SED
and time--lags among variations at different frequencies, 
are significantly and systematically larger than those 
suggested by radio luminosity
data (Dondi \& Ghisellini \cite{dondi}, Ghisellini et al. \cite{gg98},
Tavecchio et al. \cite{taold}).

Thanks to the Hubble Space Telescope (HST), faint optical
nuclear components have been recently detected in FR~I galaxies
(Chiaberge et al. \cite{paperI}). 
A strong linear correlation is found between this optical and 
the radio core emission which strongly argues 
for a common non-thermal origin. This suggests that the optical cores 
can be identified with synchrotron radiation produced in a 
relativistic jet, qualitatively supporting the unifying model for FR~I 
and BL Lacs. 

These information offer a new possibility of
verifying the unification scheme, by directly comparing the properties
of the optical and  radio cores
of radio galaxies with their putative aligned (beamed) counterparts,
analogously to the procedure followed for the radio cores.
X-ray observations also provide useful constraints to the nuclear 
emission of FR~I sources (e.g. Hardcastle \& Worrall \cite{hard99}).

The main advantage of using
multifrequency data is the possibility of directly comparing the full
broad band spectral distributions of these two classes of sources and
eventually shed light on the apparent discrepancy in the Lorentz
factors inferred from different approaches.

The paper is organized as follows.  The (complete) samples of BL Lacs
and radio galaxies are presented in Sect. \ref{samples}.  In
Sect. \ref{corext} we compare separately the core radio and optical
emission of beamed and unbeamed objects with similar extended radio
power. From this we infer the Lorentz factors requested by the
unification scheme within the simplest scenario in which the radiation
is emitted by a single uniform region of the relativistic jet.
In Sect. \ref{lrlo} the radio and optical data are considered together
and, starting from the observed SED of BL Lacs, we derive the expected
properties of the nuclear emission of FR~I, by taking into account the
spectral dependence of the relativistic transformations. As the
single--region picture does not account for the observed properties,
in Sect. \ref{multijet} we explore a (simple) alternative scenario and
test it also against the X-ray information.  Summary and conclusions
are presented in Sect. \ref{summary}.

\section{The samples}
\label{samples}
\subsection{FR~I radio galaxies}

Our complete sample of radio galaxies comprises all the FR~I sources
belonging to the 3CR catalogue (Spinrad et al. \cite{spinrad}),
morphologically identified as FR~I.  The redshifts of these objects
span the range $z = 0.0037$ -- $0.29$, with a median value of $z=
0.03$, and the total radio luminosities at 1.4 GHz are between
$10^{30.2}$ and 10$^{34.2}$ erg s$^{-1}$ Hz$^{-1}$ ($H_0= 75$ km
s$^{-1}$ Mpc$^{-1}$ and $q_0=0.5$ are adopted hereafter).  We exclude
from the original sample the peculiar object \object{3C~386}, as
discussed in Paper I.  The optical and radio data are from Paper~I,
while the X-ray ones from Hardcastle \& Worral (\cite{hard99}). The
optical data are extrapolated to the V band using a spectral index
$\alpha_o=1$ ($F_\nu \propto \nu^{-\alpha}$).

\subsection{Radio and X-ray selected BL Lacs samples}

We consider both the complete sample of 34 radio selected BL Lacs
derived from the 1Jy catalog (Stickel et al. \cite{stickel91}, K\"uhr
et al. \cite{kuhr}), and the BL Lac sample selected from the
{\it Einstein} Slew survey (Elvis et al.  \cite{elvis}, Perlman et
al. \cite{perl96}), which comprises 48 objects, and it is nearly
complete.  The extended radio power (at 1.4 GHz) $L_{ext}$ [erg
s$^{-1}$ Hz$^{-1}$] spans the ranges $10^{30.1}-10^{33.8}$ (1Jy BL Lacs, 
Kollgaard et al. \cite{koll96}) and
$10^{29.1}-10^{33.4}$ (Slew survey BL Lacs, 
Kollgaard et al. \cite{koll96}, Perlman et al. \cite{perl96}); 
the redshifts are between 0.049
and 1.048 (median $z = 0.501$) and between 0.031 and 0.513
(median $z = 0.188$) for the two samples, respectively (the redshifts
of the Slew BL Lacs are taken from the data collected by Fossati et al. 
\cite{gfos}).

Instead of classifying BL Lacs according to their selection spectral
band, in the following we adopt the definitions of high and low
energy peaked BL Lacs (HBL and LBL respectively), which are based on
the position of the (synchrotron) emission peak in the spectrum and
therefore more indicative of the physical characteristics of the
objects (Giommi \& Padovani \cite{giopad94}, Fossati et al.
\cite{gfos}). Of the 34 objects belonging to the 1 Jy sample, 32 are
classified as LBL and 2 as HBL, while of the 48 X-ray selected BL
Lacs, 40 are HBL and 8 are LBL.

The spectral data for both samples of BL Lacs are taken from Fossati
et al. \cite{gfos}.

\section{Core versus extended luminosity}
\label{corext}

\begin{figure}
\resizebox{\hsize}{!}{\includegraphics{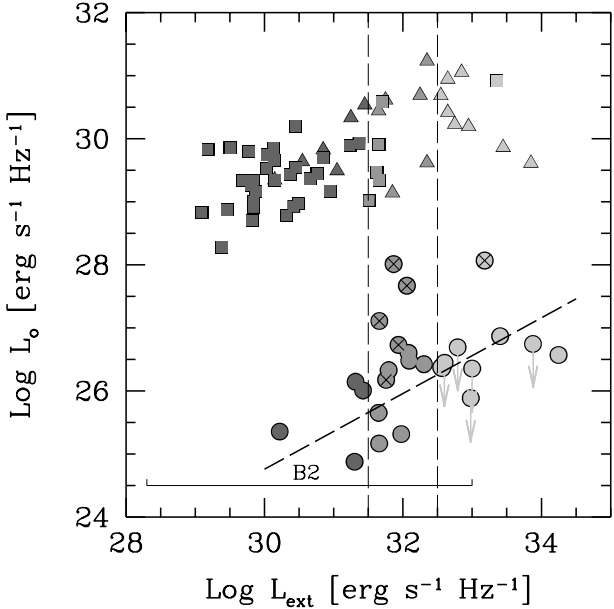}}
\caption{Optical core luminosity (V band) versus radio extended
luminosity at 1.4 GHz for FR~I (circles), 
HBLs (squares) and LBLs (triangles). 
The grey scale refers to the three bins of extended radio power. 
The dashed line represents the linear fit to the FR~I sample,
having excluded the most aligned sources, here marked with crosses
(see text). The range of extended power covered by the B2 sample of
radio galaxies is also indicated.}
\label{ext_o}
\end{figure}

\begin{figure}
\resizebox{\hsize}{!}{\includegraphics{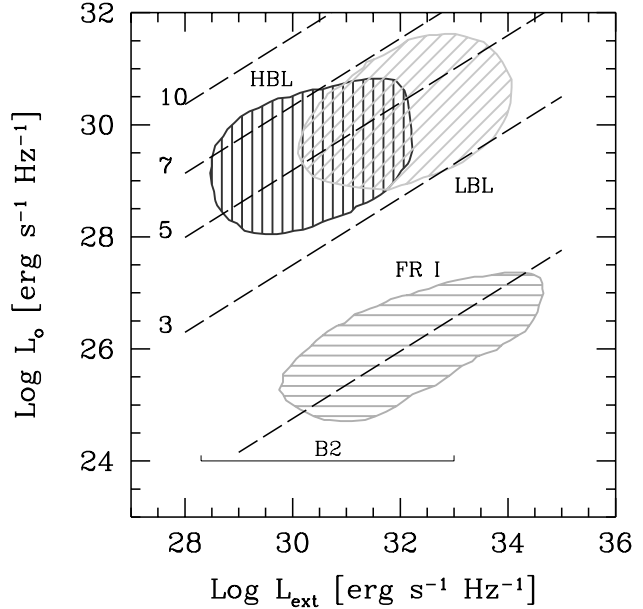}}
\caption{The regions occupied by the three samples in the optical
luminosity versus extended radio luminosity plane, as for
Fig. \ref{ext_o}. 
The dashed lines indicate the correlation found
between these two quantities when shifted by beaming effects for the
values of the bulk Lorentz factor marked on the left.}
\label{lext_lo_beam}
\end{figure}

According to the unification models the beamed and unbeamed
populations must cover the same range of extended luminosity, as this
is considered to be isotropic. On the contrary, emission from
the core is affected by beaming: radio galaxies should have a fainter
central component, whose intensity would depend on the Doppler factor
$\delta=[\Gamma(1-\beta\cos\theta)]^{-1}$, where
$\Gamma=(1-\beta^2)^{-1/2}$, $\beta c$ is the bulk velocity of the
emitting plasma and $\theta$ the angle between the direction of the
jet and the line of sight.
The transformation law for the specific flux density is
in fact $F_{\nu}=\delta^{p+\alpha} F^{\prime}_{\nu}$, where the
primed quantity refers to the comoving frame, $\alpha$ is the local
spectral index, $p=2$ for a continuous jet and $p=3$ for 
a moving sphere.

Therefore the comparison of the core emission of beamed objects and
their parent population with similar extended emission provides a
direct estimate of the Lorentz factor of the radiating plasma, if the
typical observing angles are known. With this aim and similarly to
what has been done in the radio band (e.g. Kollgaard et al.
\cite{koll96}), we plot the optical V band luminosity ($L_o$) vs the
extended radio luminosity at 1.4 GHz ($L_{ext}$) for the three samples
(Fig. \ref{ext_o}).  

First we should note that the HBL objects do not fully share the
range of extended radio power of the 3CR radio galaxies (the HBL total
luminosities are in fact more similar to the objects belonging to the
B2 sample of low power radio galaxies).  Conversely, $L_{ext}$ of LBL
well match those FR~I of the 3CR catalog.

Also notice
that the regions occupied by the two samples of BL Lacs appear to be
continuously connected, the lower radio power BL Lacs (which are HBLs)
and the higher radio power ones (LBLs) having an optical luminosity
which weakly increases for increasing extended luminosity.  Because of
this trend, in order to compare sources with the same $L_{ext}$ we
have sub-divided the samples into three bins, namely: $\log L_{ext}$
[erg s$^{-1}$ Hz$^{-1}$] $< 31.5$, $\log L_{ext}$ between 31.5 and
32.5, and $\log L_{ext} > 32.5$.  

We thus calculate the median values of the observed nuclear luminosity
of FR~I and BL Lacs in each interval of extended power. 
BL Lacs are on average 4 orders of magnitude
brighter than FR~I cores. We can assume
that BL Lacs are observed 
\footnote{At the particular angle
$\theta\sim 1/\Gamma$, one obtains $\delta=\Gamma$.} 
at $\theta\sim 1/\Gamma$ and FR~Is at $\theta=60^{\circ}$: 
in fact, for an isotropic
distribution of objects, $\theta=60^{\circ}$ corresponds to the median
angle if, as it is in the case of FR~I, the scatter in the optical
luminosity is dominated by relativistic beaming.  
Bulk Lorentz factors
$\Gamma$ $\sim 4$ for the case of an emitting
sphere and $\sim 6$ for a continuous jet are required in order to
account for the different core luminosities of FR~I and BL Lacs in
each bin of extended power. An optical spectral index $\alpha_o=1$ is
assumed for all sources (independent of beaming).

An alternative method to estimate $\Gamma$ relies on the fact that,
for a randomly oriented sample, the best fit regression line of a
luminosity distribution corresponds to the behavior of sources
observed at $\sim 60^{\circ}$, once the most core dominated objects
are excluded (Kollgaard et al. \cite{koll96}).  We thus determine the
best fit regression of FR~I in the $L_{ext}-L_{o}$ plane, after
excluding from the sample the 5 objects in which optical jets are
detected. These sources, in fact, have the most luminous optical
cores, are among the most core dominated objects in the radio band,
and their radio jets are shorter, indicating that they are pointing towards
the observer (Sparks et al. \cite{sparks}).  Interestingly, we obtain
that there is a remarkable correlation ($P > 99.9 \%$) between $\log
L_{ext}$ and $\log L_o$, among the remaining 20 
``highly misoriented''
objects, although with a slope ($\sim 0.6$) marginally steeper than
the correlation between $L_{ext}$ and core radio luminosity ($L_r$)
found by Giovannini et al. (\cite{gg2}) for a larger sample of radio
galaxies.  In Fig. \ref{lext_lo_beam} we show the regions in which the
three samples are located in the $L_{ext}-L_o$ plane, and the dashed
lines represent the ``beamed'' FR~I population as observed under an
angle $\sim 1/\Gamma$ in the case of $p=3$.  Also with this method
$\Gamma$ $\sim 5$ ($\sim 7$) are required to displace the FR~I to the
regions occupied by both HBL and LBL for $p=3$ ($p=2$).

\begin{figure}
\resizebox{\hsize}{!}{\includegraphics{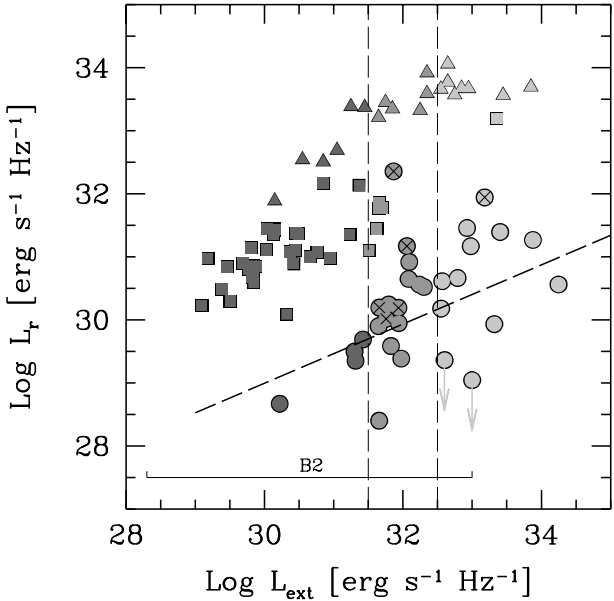}}
\caption{Radio core luminosity (at 5 GHz) versus radio extended
luminosity at 1.4 GHz for FR~I (circles), X-ray selected (squares) and
radio-selected BL Lacs (triangles). 
The dashed line is the correlation between these two
quantities found for a larger sample of galaxies by Giovannini et al.
(\cite{gg2}) and converted to 1.4 GHz using $\alpha_r=0.7$.}
\label{extcore}
\end{figure}

Let us now consider $L_r$ (at 5 GHz) versus $L_{ext}$
(Fig. \ref{extcore}), analogously to what is shown by Kollgaard et
al. (\cite{koll96}) for a larger sample of radio galaxies (which also
includes our objects). The typical radio core luminosities of HBL and
LBL are significantly different, the latter objects being on average
about one order of magnitude more luminous than the former ones.
Conversely, as we have already pointed out, no substantial difference
between the two classes is found in the case of $L_o$.

These results have been initially attributed to a
different amount of beaming for X-ray and radio-selected BL Lacs
(i.e. different angle of
sight and/or different jet velocities\footnote{If so the inferred
Lorentz factors, relative to our sample,
are $\Gamma=4$ (3) for HBL $\Gamma=10$ (5) for LBL, in the case of 
p=2 (p=3)}) while more recently a
consistent picture has emerged where this diversity can be accounted
for by the different shape of their intrinsic SED (e.g. Padovani
\cite{pado92}, Ghisellini \& Maraschi \cite{ggma}, Giommi \& Padovani
\cite{giopad94}, Fossati et al. \cite{gfos}).  
The role of these two
scenarios will be further explored in the next section, through the
comparison of the SED of both types of BL Lacs with their parents.

We conclude that the Lorentz factors inferred from the comparison of
the radio, but also optical
emission of FR~I and BL Lacs, are consistent with those
previously estimated from the statistics of these sources within
the unifying scheme.
However, as already mentioned, such values are significantly and 
systematically lower than those required by other independent means,
such as superluminal motions and high energy spectral constraints
(fit to the overall SED and time--lags) in both LBLs and HBLs.
(Maraschi et al. \cite{nonna92}, Sikora et al. \cite{sikora},
Celotti et al. \cite{mkntev}, Tavecchio et al. \cite{taold}). 
These latter methods require a value of the Doppler factor
$\delta$ in the range 15--20 for the region emitting most of the 
radiation in both HBLs and LBLs.
The need for high degrees of beaming will constitute a crucial point 
in the following.

\section{FR~I and BL Lac in the  $L_o$ and $L_r$ plane}
\label{lrlo}

\begin{figure} 
\resizebox{\hsize}{!}{\includegraphics{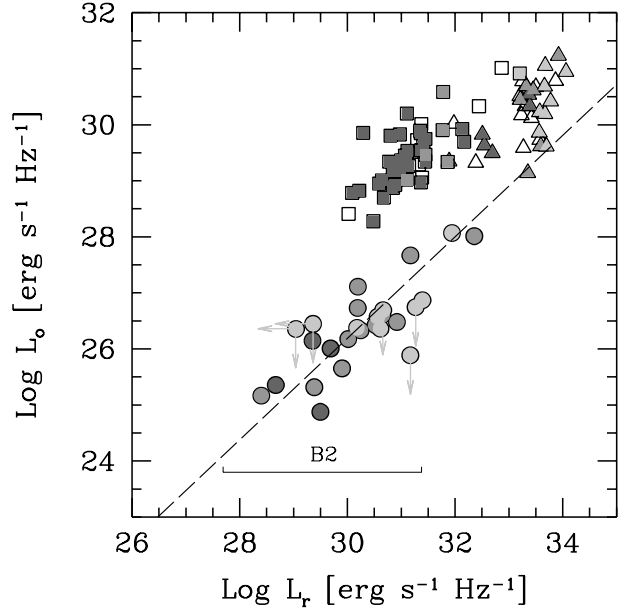}}
\caption{BL Lacs and FR~I radio galaxies in the $L_r-L_o$ plane.
Empty symbols are objects with no data on their extended radio power,
filled symbols and grey scale are as in Fig. \ref{ext_o}.  The dashed
line is the radio-optical core correlation (Paper~I). The range of
core luminosity of the B2 radio galaxies is also reported.  Notice
that the B2 cores could at most extend this correlation by $\sim 1$
order of magnitude towards lower luminosities.}
\label{lumbl} 
\end{figure}

Since for the first time multifrequency data are available also for the
nucleus of radio galaxies, we can now directly compare the spectral
properties of beamed objects and their parent population: this new
approach can thus combine information from the SED of BL Lacs with those
inferred from their relation with FR~I.
In particular in Paper I and in Chiaberge et al. \cite{papII} we showed how 
the location of sources in the  $L_o$ and $L_r$ plane
represents a very useful tool to discuss their nuclear properties.
In Fig. \ref{lumbl} we show the optical vs radio core luminosity
for the three samples. 
The dashed line represents the 
(almost linear) correlation found between these two quantities among 
the FR~I sources (Paper~I).  
Radio galaxies, HBL and LBL occupy different
regions of this plane: LBLs are located only marginally above
the continuation of this correlation, while HBL are $\sim
2$ order of magnitude brighter in the optical with respect to other
objects for a given radio luminosity.

In order to determine how beaming affects the observed luminosities
and thus how objects could be connected in this plane, we
consider the SED of BL Lacs, observationally much better determined,
and calculate the observed spectrum of the misaligned objects, by
taking into account relativistic transformations.

In fact, an important point, previously neglected, is that these
transformations depend on the spectral index in the band considered,
which in itself might change as a function of the degree of beaming.
Therefore, in order to correctly de--beam the SED of BL Lacs, a
continuous representation of it and an estimate of the bulk Lorentz
factor of the emitting region are needed (it is again assumed
$\theta=1/\Gamma$). 
While any continuous description of the SED and
typical Lorentz factors can be used, we derive both of them by adopting a
homogeneous synchrotron self--Compton emission model to reproduce the
observed SEDs (e.g. Chiaberge \& Ghisellini \cite{cg},
Ghisellini et al. \cite{gg98}, Mastichiadis \& Kirk \cite{masti}).
This approach has the advantage 
of considering both the emission and dynamical
($\Gamma$) properties self--consistently. 
\footnote{These models, in fact, satisfy time variability constraints,
assume continuous injection of particles, radiative and adiabatic cooling, 
$\gamma$--$\gamma$ collisions and pair production.
The continuous curves shown in Fig. 5 are then not only 
interpolating curves, but physically possible fits to the data.}
The bulk velocities obtained in this way are fully compatible with 
those inferred from the already mentioned constraints (Sect. \ref{corext}).

\begin{figure} 
\resizebox{\hsize}{!}{\includegraphics{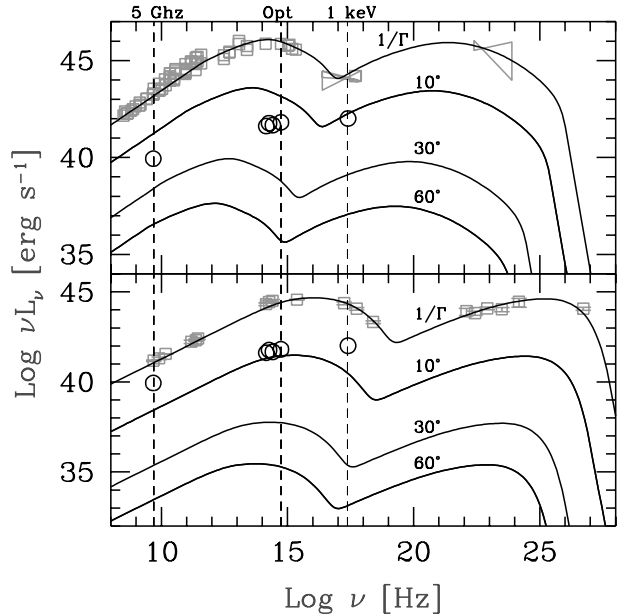}}
\caption{Spectral energy distributions of Mkn 421 (lower panel) and
PKS 0735+178 (top panel) and debeamed SED for different viewing angles
in the case of a single emitting component. The Lorentz factors
inferred for the two sources are $\Gamma= 20$ and $\Gamma=16$, respectively. 
For comparison, we report
(empty circles) the radio, IR and optical (HST) and X-ray (ROSAT) data
for \object{3C~264} (Capetti et al. \cite{hstx}).  The
(non-simultaneous) data for PKS 0735+178 are taken from the literature
(NED). The (quasi-simultaneous) data for Mkn 421 are from Macomb et
al. (\cite{macomb}).}
\label{shiftsed}
\end{figure}

\begin{figure} 
\resizebox{\hsize}{!}{\includegraphics{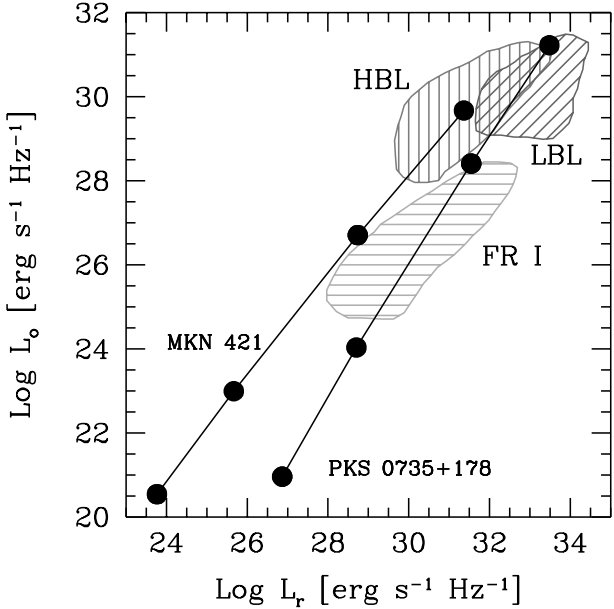}}
\caption{Debeaming trails in the radio-optical luminosity plane for
Mkn 421 and PKS 0735+178, in the frame of a single emitting region
model. The filled circles correspond to the predicted 
luminosities of objects at different viewing angles. Top to
bottom: $\theta= 1/\Gamma$, 10$^\circ$, 30$^\circ$ and 60$^\circ$.}
\label{mkn1}
\end{figure}

Let us firstly consider single objects for which the SED is well sampled, 
namely \object{Mkn~421} (a typical HBL) and \object{PKS 0735+178} 
(a typical LBL).
We model the observed SED as explained, derive the value of 
$\Gamma$($=\delta$) for the two sources and then calculate their 
corresponding observed SEDs for different orientations.  
Clearly the net effect of debeaming is a ``shift'' of the SED towards 
lower luminosities and energies (see Fig. \ref{shiftsed}).

Notice that, as the model is appropriate for the optically thin part of
the spectrum, in order to account for the radio emission, which
necessarily has to be produced on larger scales, we linearly
extrapolate the fit from the infrared-mm spectral region.  However, at
an angle of $\theta=60^{\circ}$ and for the Lorentz factors derived
from the model, $\Gamma \sim 15-20$, the observed (debeamed) radiation
at 5 GHz corresponds to what is seen in BL Lacs at far infrared
frequencies (respectively $\sim 500 - 300 \mu m$, see
Fig. \ref{shiftsed}) and therefore the debeamed points in
Fig. \ref{mkn1} represent the correct predicted luminosities of the BL
Lac component at 5 GHz.

The resulting debeamed optical and radio luminosities are reproduced
in Fig. \ref{mkn1}.  The dash--dotted lines represent ``debeaming
trails'' and the filled circles the calculated debeamed luminosities
for $\theta=1/\Gamma$ (i.e. the BL Lac itself), $10^{\circ}$,
$30^{\circ}$ and $60^{\circ}$.  Most noticeably, for
$\theta=60^{\circ}$ -- which is the mean angle of sight for the
misaligned population -- the BL Lac component is about four orders of
magnitude below the radio galaxy region in the optical, and two/four
in the radio band. 

While equivalently incompatible with the FR~I population, the HBL and
LBL move on different trails.  This is due to the different shape of
their SED (see Fig.  \ref{shiftsed}), and in particular to the
position of the synchrotron peak frequency: if -- for increasing
values of $\theta$ -- in the rest frame the peak overcomes the optical
band, the spectral index steepens and the optical flux drops more
rapidly than the radio one.

\begin{figure} 
\resizebox{\hsize}{!}{\includegraphics{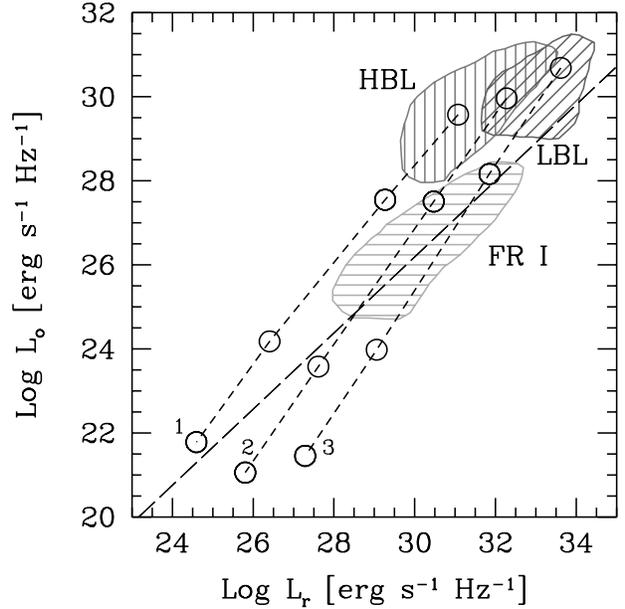}}
\caption{Debeaming trails in the radio-optical luminosity plane for
average BL Lacs SED, in the frame of a single emitting component
model.  The filled circles correspond to the predicted optical-radio
luminosity for different angles of sight (top to bottom:
$\theta=1/\Gamma$, 10$^\circ$, 30$^\circ$ and 60$^\circ$).}
\label{sedmed}
\end{figure}

Another remarkable result is that the debeaming trail of the HBL does not
even cross the region occupied by radio galaxies in the $L_r-L_o$
plane. As this might be a serious problem for the unified scheme, we
further examine this issue. In particular we closely examine the
effect of the spectral shape and its relation with the intrinsic
luminosity by considering three different SED, which represent the
whole family of BL Lacs, from HBLs to LBLs. 
\footnote{The three SED
correspond to different bins of radio luminosity (at 5 GHz) -- which
appears correlated with the bolometric luminosity and the position of
the peak frequency (Fossati et al. \cite{gfos}, Ghisellini et
al. \cite{gg98}).} 
In Fig. \ref{sedmed} we plot the resulting
trails: once again, as in the cases of Mkn 421 and PKS 0735+178, the
expected nuclear luminosity is 10-10$^4$ times fainter than what
observed in FR~I and the debeaming trail for the lower luminosity
object (a typical HBL) does not even cross the FR~I region.  
Note that if the luminosity is indeed related to the shape of the SED, 
this discrepancy would exacerbate for even fainter BL Lacs.  
In fact for
HBL the radio and optical spectral slopes can be considered constant
as the viewing angle increases, resulting in a linear (one to one)
debeaming trail, parallel to the FR~I correlation.

Summarizing: the radio and optical luminosities of BL Lacs and FR~I are
not consistent with the simplest predictions of the unifying scheme, if a
single emitting region is responsible for the different broad band
spectral properties of the beamed and parent populations.  More
specifically: i) Lorentz factors $\sim 15$, as derived from the high
energy spectral properties, underestimate the predicted emission from the
parent population; ii) the relative ratio of radio to optical luminosity
of HBL is inconsistent with the observed FR~I spectra.  In the next
section we discuss and test a possible solution to these discrepancies.

\section{A jet velocity structure}
\label{multijet}

\begin{figure} 
\resizebox{\hsize}{!}{\includegraphics{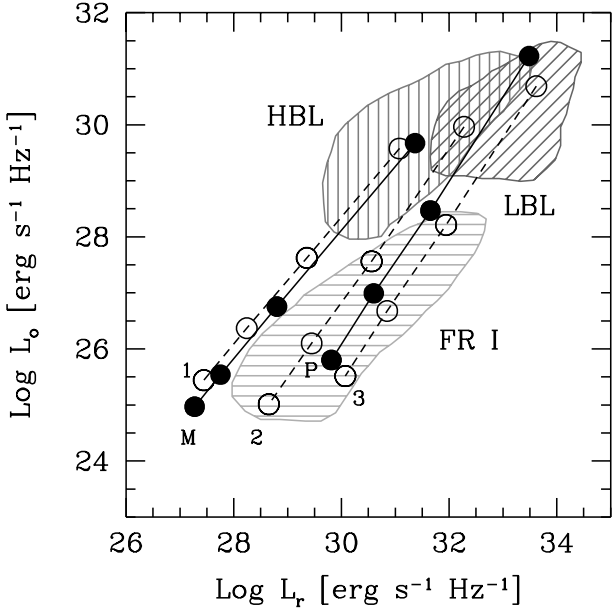}}
\caption{Debeaming trails in the radio-optical luminosity plane, for
the case of the two-velocity jet. The five curves correspond to
average SED (1,2 and 3), Mkn 421 (M) and PKS 0735+178 (P). For the
average SEDs $\Gamma_{spine}=15$ and $\Gamma_{layer}=2$; for Mkn 421
$\Gamma_{spine}=20$ and $\Gamma_{layer}=1.2$; for PKS 0735+178
$\Gamma_{spine}=16$ and $\Gamma_{layer}=1.5$. 
Circles correspond to the predicted optical-radio
luminosity for different angles of sight (top to bottom:
$\theta=1/\Gamma$, 10$^\circ$, 30$^\circ$ and 60$^\circ$).}
\label{mkn2} 
\end{figure}

We have shown that the high bulk Lorentz factors required by the
emission models of BL Lacs imply that if such objects are observed at
$60^\circ$ the resulting spectral properties are not compatible with
what is observed in the nuclei of radio galaxies. 
And indeed the
previous comparison of the core emission of FR~I and BL Lacs (see
Sect. \ref{corext}) led to lower values of $\Gamma$.  
How can these results be reconciled within the unifying scenario?  
A possible and
plausible effect, which could account for this discrepancy, is
provided by the existence of a distribution in the bulk velocity of
the flow, with the emission from plasma moving at different speeds
dominating the flux observed at different viewing angles.

Let us consider this hypothesis in the frame of the unification scheme
and examine the simplest case, i.e. a model with two axisymmetric
components having the same intrinsic luminosity and spectra. 
In other words, the
only difference between the center and the layer of the jet is the
bulk Lorentz factor which is determined for the {\it spine}
($\Gamma_{spine}$) by modeling the BL Lac SED, while for the {\it layer}
($\Gamma_{layer}$) by requiring that the debeamed BL Lac match the 
FR~I distributions in the $L_{ext}-L_o$ and $L_r-L_o$ planes.

The monochromatic intensity emitted by the jet is therefore calculated as

\begin{displaymath}
I(\nu,\theta) = \delta^3_{spine}(\theta)
                I^{\prime}(\nu/\delta_{spine})+
                \delta^3_{layer}(\theta)
                I^{\prime}(\nu/\delta_{layer}), 
\end{displaymath}
where 
$I^{\prime}$ is the comoving intensity.

\begin{figure}
\resizebox{\hsize}{!}{\includegraphics{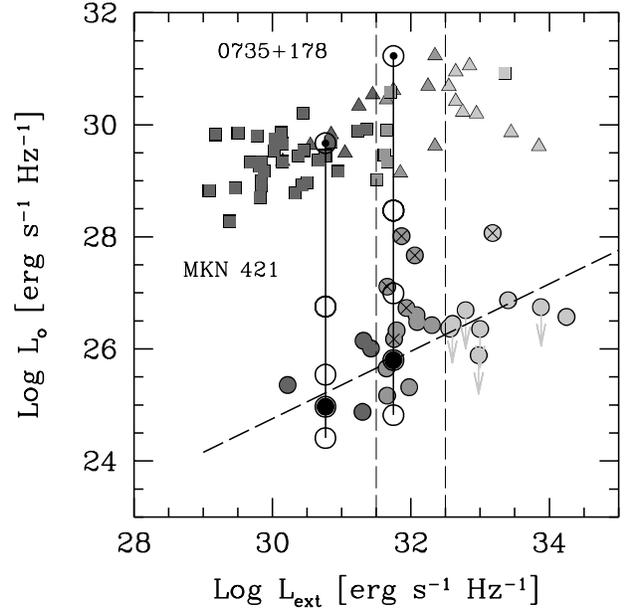}}
\caption{Debeaming trails for the optical emission of 
Mkn~421 and PKS~0735+178 in the two-velocity 
jet scenario. The circles connected by the vertical lines
correspond to the sources observed at angles of
(top to bottom) $1/\Gamma$, 10$^\circ$, 30$^\circ$, 60$^\circ$ 
(filled black circles) and 90$^\circ$. The values of $\Gamma_{layer}$ (1.2 for 
and 1.5 for Mkn~421 and PKS~0735+178, respectively) are chosen
in order for the luminosity at 60$^\circ$ to correspond to the median 
value for each bin of extended radio power.}
\label{debext} 
\end{figure}

The predicted luminosity trails for the two specific BL Lacs 
are shown in Fig. \ref{mkn2}. Values of
$\Gamma_{layer}$ are set to 1.2 and 1.5 for Mkn 421 and PKS 0735+178
respectively, so that the point of each
trail corresponding to the angle $\theta=60^{\circ}$ falls approximatively
onto the median of the FR~I optical core luminosity in each $L_{ext}$ bin 
(Fig. \ref{debext}).

\begin{figure}
\resizebox{\hsize}{!}{\includegraphics{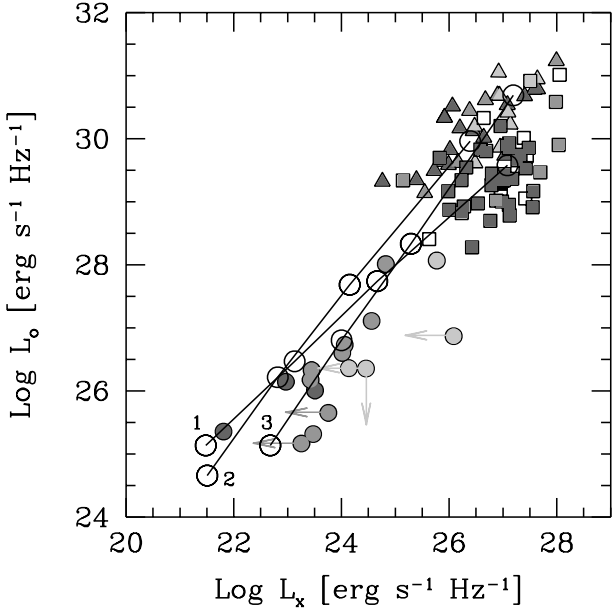}}
\caption{Debeaming trails in the optical-X-ray luminosity plane for
the average BL Lacs SED. Two-velocity jet with $\Gamma_{spine}=15$ and
$\Gamma_{layer}=2$ for all SED. Circles are FR~I, squares and triangles
represent X-ray and radio selected BL Lacs, respectively. 
Grey scales correspond to bins of extended
radio luminosity, as for Fig. \ref{ext_o}.  Open triangles and squares
are sources without measurement of their extended radio power.
Open circles correspond to the predicted optical-X-ray
luminosity for different angles of sight (top to bottom:
$\theta=1/\Gamma$, 10$^\circ$, 30$^\circ$ and 60$^\circ$).}
\label{lumox} 
\end{figure}

By using the same value of $\Gamma_{layer}$ 
this two-velocity model satisfactorily
predicts the properties of the debeamed counterparts of LBL and in particular
it reproduces the FR~I location in the $L_r-L_o$ plane.

Conversely this picture can not account for the
observed optical--radio properties of debeamed HBL.
The luminosity of objects seen at $\theta =
60^{\circ}$ is close to the lower limit of the FR~I region in the
optical, but still one order of magnitude fainter in the radio. 
We must stress however that
the extended radio powers of HBLs correspond more closely to
the range covered by the B2 radio galaxies.
Clearly, any firm statement on this issue must await for the
analysis of the nuclear properties of the B2 sample,
but the extrapolation
of the 3CR radio-optical correlation does not match the
debeamed predicted luminosities of HBL.
This result does not depend on the specific value of $\Gamma_{layer}$
adopted since, as already discussed in Sect. \ref{lrlo}, 
the HBL trails run almost parallel to the radio-galaxies locus.

A further modification of this model is thus required for the HBL unification.
Without altering the comoving spectra of the two components, 
the simplest change is to assume a lower Doppler factor for {\it spine} 
in the radio emitting region, as might
be the case if the flow slows down between the optical and the radio
emitting sites. This would increase the initial 
slope of the debeaming trail which would rapidly 
reach the FR~I region. 

\subsection{Constraints from the X--ray observations}

The limited angular resolution makes the analysis of X-ray observations of FR~I
sources less straightforward than in the radio and optical bands.
In particular it is necessary to disentangle any non thermal
nuclear radiation from the often dominant emission of the hot gas
associated with the galactic corona and/or galaxy cluster.
Nonetheless they provide useful constraints to high energy 
nuclear emission of these radio-galaxies.

We can test the validity of the two-velocity jet scenario
by considering also this X-ray emission. In Fig. \ref{lumox} we report the
debeaming trails in the $L_X - L_o$ plane 
for the average SEDs assuming the same jet parameters as
before: the predicted powers in both bands appear to be
consistent with the observed properties of radio galaxies,
supporting the presence of a less beamed plasma
component (layer) dominating the emission in the parent population
also in the X-ray band.
Conversely, there is no need for a different amount of beaming in these two
bands. 
This is somehow reassuring, as 
optical and X--ray are believed to originate co--spatially.

In addition, X--ray data can be used to define the location of radio--galaxies
also in the broad band spectral indices plane. 
The spectral characteristics of blazars are often
represented in the plane defined by 
$\alpha_{ro}$ (5GHz-5500\AA) and $\alpha_{ox}$(5500\AA-1 keV).
It is therefore worthwhile to determine how relativistic beaming
affects the position of the objects also in this plane.  While an
approximated relation between the BL Lacs and FR~I broad band spectral
slopes is derived in \ref{formula} in the case of constant local
spectral indices, changes in the local spectral slopes are properly
taken into account.

\begin{figure}
\resizebox{\hsize}{!}{\includegraphics{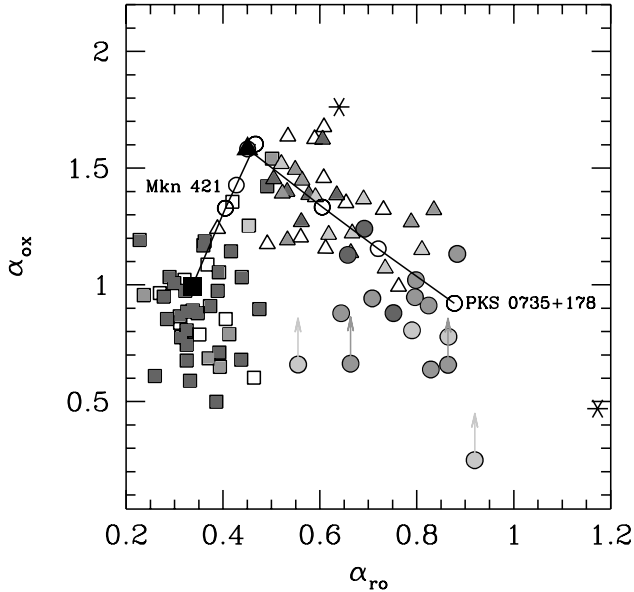}}
\caption{Debeaming trails in the $\alpha_{ro}-\alpha_{ox}$ plane for
Mkn 421 and PKS 0735+178. The jet parameters are the same as in
Fig. \ref{mkn2}. The black filled symbols correspond to the observed
BL Lac ($\theta=1/\Gamma$), while the empty circles represent the
predicted position for different angles of sight ($\theta=10^\circ$,
30$^\circ$ and 60$^\circ$) in the frame of a two-velocity jet model.
The two asterisks indicate sources observed at $\theta=60^\circ$
in the case of a single emitting component.}
\label{alxro} 
\end{figure}

In this plane (Fig. \ref{alxro}), as already well established, HBL and
LBL occupy the left (i.e. flatter $\alpha_{ro}$) and the top-center
(i.e. steeper $\alpha_{ox}$) regions, respectively and their different
position is accounted for by their different SEDs (e.g. Fossati et
al. \cite{gfos}), i.e. reflects the position of the peak of the
emission.  The FR~I region is instead well defined at the center of
the diagram.  The debeaming trails for Mkn 421 and PKS 0735+178 are
also shown
\footnote{Note that again the two trails differ because of the
different SEDs: for Mkn 421 the peak of the synchrotron component is
between the optical and the X-ray bands and therefore the effect of
debeaming is, initially (i.e. for increasing $\theta$) to steepen both
$\alpha_{ro}$ and $\alpha_{ox}$; when the Compton peak enters the
X-ray band, $\alpha_{ox}$ flattens.  Instead, for PKS 0735+178 the
synchrotron peaks between the radio and the optical, and the 1 keV
flux is due to Compton and, for the range of angles of sight
considered here, the change in the spectral indices turns out to be
monotonous.}: the empty circles correspond to $\theta = 1/\Gamma$,
$10^{\circ}$ $30^{\circ}$, $60^{\circ}$ in the case of the two
components model, while the two asterisks represent each source as
observed at $60^{\circ}$ in the case of a single emitting region. 
PKS 0735+178 falls in the radio galaxy region for
$\Gamma_{layer}=3$ or less, while Mkn 421 does not intersect this area
either in the single or in two component models, confirming the results of the 
analysis presented above.

\section{Summary and conclusions}
\label{summary}

With the aim of exploring the viability of the unification scenario
between (HBL, LBL) BL Lacs and FR~I radio galaxies we have compared
their nuclear emission in the radio, optical and X--ray bands.

We have firstly considered these spectral regions separately,
comparing the nuclear emission of the two classes of objects for
similar extended radio power.  As the core radiation of BL Lacs is
enhanced by relativistic beaming, we derived the bulk Lorentz factors
requested to account for the observed distribution.  The values of
$\Gamma$ thus inferred are not compatible with the higher bulk
velocities requested by theoretical arguments, such as the pair
production opacity and the spectral modeling of the SED of BL Lacs.

We then examined the core emission of three samples in the $L_r-L_o$
plane. In the frame of the simplest one-zone emission model, we
calculated debeaming trails of the BL Lac broad band emission as
predicted by the relativistic transformation for an increasing angle
of sight.  We found that the model does not account for the observed
spectral properties of FR~I, as expected from the above inconsistency
of the Lorentz factors.

The simplest and rather plausible hypothesis to account for this
discrepancy within the unification scenario is to assume a structure
in the jet velocity field, in which a fast spine is surrounded by a
slow layer. Note however that the slower jet component must be
relativistic in order to explain the anisotropic radiation of radio
galaxy cores (e.g. Capetti \& Celotti \cite{ac2}).  The observed flux
is dominated by the emission from either the spine or the slower
layer, in the case of aligned and misaligned objects, respectively.

Interestingly, the existence of velocity structures in the jet has
been suggested by various authors (Komissarov \cite{komiss}, Laing
\cite{laing93}) in order to explain some observed properties of FR~I
(and FR~II) jets, such as the structure of the magnetic field in FR~I
which appears to be longitudinal close to the jet axis and transverse
at the edges. Swain et al. \cite{swain} obtained VLA images of
\object{3C~353} (an FR~II with straight jets), finding that a model
consisting in a fast relativistic spine ($\beta >0.8$) plus a slower
outer layer ($\beta <0.5$, but still relativistic in order to produce
the observed jet-counterjet intensity asymmetry) could account for the
apparently lower emissivity near the jet axis. 
Similar behaviours have been inferred for the two low luminosity
radio galaxies \object{M~87} (Owen et al. \cite{owenm87}) and 
\object{B2~1144+35} (Giovannini et al. \cite{gg299}).
Furthermore, Laing et
al. (\cite{laing99}) showed that the jet asymmetries in FR~I can be
explained by means of a two-speed model. As a consequence, they argued
that the lower velocity component dominates in the cores of the
edge-on sources, while the fast spine emission dominates the end-on
ones.  
This possibility might be also supported by recent numerical
simulations of relativistic jets (Aloy et al. \cite{aloy}).

The same indication has been found through different approaches.
Capetti \& Celotti \cite{ac2} reveal a trend in the radio galaxy/BL
Lac relative powers with the line of sight, which is consistent with a
slower (less beamed) component dominating at the largest angles.
Capetti et al. \cite{hstx} consider the same issue by examining
the more detailed SED of five radio galaxies and consider their beamed
counterparts. They found that while the spectral shapes of
\object{3C~264} and \object{3C~270} can be reconducted to those of BL
Lacs, the required ratio of beaming factors, i.e. $\delta_{\rm BL
Lac}/ \delta_{\rm FR~I}\sim 10-100$, implies that the corresponding BL
Lacs would be overluminous. The inclusion of a slower (less beamed)
jet component seems to be a plausible explanation.

We found that Lorentz factors of the layer $\Gamma_{layer}\sim 2$ 
can account for
the unification of FR~I (of the 3CR) with LBL and intermediate
luminosity BL Lacs.  Instead the debeaming trails for the lowest
luminosity HBL do not cross the FR~I region in the $L_r-L_o$
plane. While the HBL behavior should be compared with that of radio
galaxies with which they share the extended radio power (e.g. those of
the B2 catalogue), our simple two-component jet model could not
account for the observed properties if the cores of such low-power
FR~I radio galaxies lied on the extrapolation of the 3CR radio-optical
correlation.  The properties of such weak sources can be instead
reproduced if their radio emitting region is less beamed than the
optical one, as could be expected if the jet decelerates after the
higher energy emitting zone.

Finally, the presence of velocity structures in jets of course affects the
number counts of beamed and unbeamed sources: for example, the lack of BL Lacs 
in clusters (Owen et al. \cite{owen96}) could be attributed to 
values of typical bulk Lorentz factors higher than those derived 
from statistical arguments (Urry et al. \cite{ups91}). Intriguingly,
the very latter authors had to require a wide distribution of Lorentz
factors to account for the number densities of FR~I and BL Lacs in the
radio band.

Much has still to be understood on the dynamics and emitting
properties of relativistic jets. Multifrequency studies of the nuclear
properties of beamed sources and their parent populations and their
comparison -- according to unification scenarios which are well
supported by other independent indications -- constitute a new and
powerful tool to achieve that, both for well studied individual
sources as well as complete samples. Near IR observations by HST, mm
data and higher resolution and sensitivity by Chandra in X--rays will
further open this possibility.

Concluding, the radio, optical and X-ray nuclear emission of FR~I and
BL Lacs strongly indicate the presence of a velocity structure in the
jet if indeed these sources are intrinsically identical.  In other
words, by considering the indications of trends in the SED of blazars
emerged in the last few years (Giommi \& Padovani \cite{giopad94},
Fossati et al. \cite{gfos}) together with the constraints derived from
their unification with radio galaxies, it appears that the
phenomenology of these sources is characterized and determined by
differences {\it both} in the intrinsic SED {it and} in beaming
properties.

\begin{acknowledgements}{}
The authors thank the anonymous referee for his/her useful comments.

This research has made use of the NASA/IPAC Extragalactic Database
     (NED) which is operated by the Jet Propulsion Laboratory, California
     Institute of Technology, under contract with the National Aeronautics
     and Space Administration. 
The Italian MURST is thanked for financial support.

\end{acknowledgements}{}

\appendix

\section{Debeaming and the broad band spectral slopes}
\label{formula}

In the case of a one-component model, and under the assumption that
the local spectral indices are constant, we can derive the
transformation law for the change of the spectral slope due to
relativistic beaming.  If the flux density in the frame comoving with
the emitting region is $F^{\prime}_{\nu^{\prime}}$, the observed one
is
\begin{displaymath}
F^{\rm object}_{\nu}(\nu)=\delta^{p+\alpha}_{\rm object} 
F^{\prime}_{\nu^{\prime}}(\nu),  
\end{displaymath}
where $\delta_{\rm object}$ is the beaming factor of the same object
for different lines of sight (i.e. observed as BL Lac or as
radio galaxy).  Substituting these transformations in the definition of
the broad-band spectral index $\alpha_{12}$ (where 1 and 2 refer to
radio, optical or X-ray flux), one obtains
\begin{displaymath}
\alpha^{\rm BL Lac}_{12} = - \frac{\log\{\frac{F_2^{\rm FR~I}(\nu_2)
(\delta_{\rm BL Lac}/ \delta_{\rm FR~I})^{p+\alpha_2}}{F_1^{\rm
FR~I}(\nu_1) (\delta_{\rm BL Lac}/ \delta_{\rm
FR~I})^{p+\alpha_1}}\}}{\log(\nu_2/\nu_1)}
\end{displaymath}
which can be written as
\begin{displaymath}
\alpha^{\rm BL Lac}_{12} - \alpha^{\rm FR~I}_{12} =
	(\alpha_1-\alpha_2) \frac{\log (\delta_{\rm BL
	Lac}/\delta_{\rm FR~I})} {\log (\nu_2/\nu_1)} \, ,
\end{displaymath}
where $\alpha_1$ and $\alpha_2$ are the local spectral indices.


\begin{thebibliography}{}

\bibitem[2000]{aloy} Aloy, M. -A. , G\'omez, J. -L. , Ib\'a\~nez, J. -M. , Mart\'\i, J. -M., M\"uller, E. ,  2000, ApJ 528, L85

\bibitem[1993]{anto} Antonucci R., 1993, A\&AR, 31, 473

\bibitem[1985]{antoulve} Antonucci R. R. J., Ulvestad J. S., 1985, 
ApJ, 294, 158

\bibitem[1999]{ac2} Capetti A., Celotti A., 1999, MNRAS 304, 434

\bibitem[2000]{hstx} Capetti A., Trussoni E., Celotti A., Feretti L., 
Chiaberge M. 2000, MNRAS submitted

\bibitem[1993]{celo93} Celotti A., Maraschi L., Ghisellini G., Caccianiga A.,
Maccacaro T., 1993, ApJ, 416, 118

\bibitem[1998]{mkntev} Celotti A., Fabian A. C., Rees M. J., 1998, MNRAS 
293, 239

\bibitem[1999]{cg} Chiaberge M., Ghisellini G. , 1999, MNRAS, 306, 551

\bibitem[1999]{paperI} Chiaberge M., Capetti A., Celotti A., 1999, A\&A, 
349, 77

\bibitem[1999]{papII} Chiaberge M., Capetti A., Celotti A., 2000, A\&A,
in press

\bibitem[1995]{dondi} Dondi L., Ghisellini G., 1995, MNRAS, 273, 583

\bibitem[1992]{elvis} Elvis M. , Plummer D. , Schachter J., Fabbiano G., 
1992, ApJS, 80, 257

\bibitem[1984]{fabbiano} Fabbiano G., Trinchieri G., Elvis M., 
Miller L., Longair M. 1984, ApJ 277, 115

\bibitem[1974]{fr74} Fanaroff B. L., Riley J. M. 1974, MNRAS 167, 31

\bibitem[1998]{gfos} Fossati G., Maraschi L., Celotti A., Comastri A., 
Ghisellini G., 1998, MNRAS, 299, 433

\bibitem[1989]{ggma} Ghisellini G., Maraschi L., 1989, ApJ 340, 181

\bibitem[1993]{gg93} Ghisellini G., Padovani P., Celotti A., Maraschi L., 
1993, ApJ, 407, 65

\bibitem[1998]{gg98} Ghisellini G., Celotti A., Fossati G., Maraschi L., 
Comastri A., 1998, MNRAS, 301, 451

\bibitem[1994]{giopad94} Giommi P., Padovani P., 1994, MNRAS 268, L51

\bibitem[1988]{gg2} Giovannini G., Feretti L., Gregorini L.,
	Parma P. 1988, A\&A 199, 73

\bibitem[1999]{gg299} Giovannini, G., Taylor, G. B., Arbizzani, E. 
et al. 1999, ApJ 522, 101

\bibitem[1999]{hard99} Hardcastle, M. J., Worrall, D. M., 1999, MNRAS, 309, 969

\bibitem[1998]{henkel} Henkel C., Wang Y. P., 
Falcke H., Wilson A. S., Braatz J. A. 1998, A\&A 335, 463 

\bibitem[1992]{koll92} Kollgaard R. I., Wardle J. F. C.,  
Roberts D. H., Gabuzda D. C.  1992, AJ 104, 1687 

\bibitem[1996]{koll96} Kollgaard R. I., Palma C., Laurent-Muehleisen S. A., Feigelson E. D., 1996, ApJ 465, 115

\bibitem[1990]{komiss} Komissarov S. S., 1990, Soviet Astronomy Letters, 16, 284

\bibitem[1981]{kuhr} K\"uhr H., Witzel A., Pauliny-Toth I. I. K.,
Nauber U. 1981, A\&AS 45, 367 

\bibitem[1993]{laing93} Laing R. A., 1993, in Burgarella D., Livio M.,
O'Dea C.P., eds Space Telescope Sci. Inst. Symp. 6: Astrophysical Jets.
Cambridge University press, Cambridge, p. 95

\bibitem[1999]{laing99} Laing R. A., Parma P., de Ruiter H. R., Fanti R., 
1999, MNRAS, 306, 513

\bibitem[1996]{macomb} Macomb D. J., Akerlof C. W.,
 Aller H. D., et al., 1996, ApJl, 459, L111

\bibitem[1992]{nonna92} Maraschi L., Ghisellini G., Celotti A., 1992, 
ApJ 397, L5

\bibitem[1997]{masti} Mastichiadis A., Kirk J. G., 1997, A\&A, 320, 19

\bibitem[1993]{murphy} Murphy D.W., Browne I.W.A., Perley  
R.A. 1993, MNRAS  264  298 

\bibitem[1996]{owen96} Owen F. N., Ledlow M. J., Keel W. C. 1996, AJ 111, 53

\bibitem[1989]{owenm87} Owen F. N., Hardee P. E., Cornwell T. J., 1989, 
ApJ 340, 698

\bibitem[1992]{pado92} Padovani, P., 1992, A\&A, 256, 399

\bibitem[1996]{perl96} Perlman E. S., Stocke J. T.,
 Schachter J. F., et al., 1996, ApJS, 104, 251

\bibitem[1994]{sikora} Sikora M. , Begelman M. C., Rees M. J., 1994, ApJ 
421, 153

\bibitem[1995]{sparks} Sparks W. B., Golombek D. , Baum 
S. A. et al. 1995, ApJL 450, L55


\bibitem[1985]{spinrad} Spinrad H., Djorgovski S., Marr J., 
           Aguilar L. 1985, PASP 97, 932 

\bibitem[1991]{stickel91} Stickel M., Fried J. W., K\"uhr H., Padovani P., 
Urry C. M., 1991, ApJ 374, 431

\bibitem[1998]{swain} Swain, M. R., Bridle, A. H. Baum, S. A., 1998, ApJl, 
507, L29

\bibitem[1998]{taold} Tavecchio F., Maraschi L.,  Ghisellini G., 
1998, ApJ 509, 608

\bibitem[1999]{edo} Trussoni E., Vagnetti F.,
 Massaglia S., et al., 1999, A\&A, 348, 437

\bibitem[1989]{ulrich89} Ulrich M. H., 1989, in BL Lac Objects, ed. 
L. Maraschi, T. Maccacaro, M.H. Ulrich (Berlin, Springer), 45

\bibitem[1999]{urry99} Urry, C. M. , Falomo, R. , Scarpa, R., et al., 1999, 
ApJ, 512, 88

\bibitem[1995]{urrypad} Urry C. M., Padovani P. 1995, PASP 107, 803

\bibitem[1991]{ups91} Urry C. M., Padovani P., Stickel M. 1991, 
ApJ 382, 501

\bibitem[1994]{wb94} Worral D.M., Birkinshaw M. 1994, ApJ 427, 134

\end{thebibliography}
\end{document}